\documentclass{aastex}
\usepackage{spr-astr-addons}
\usepackage{url}

\RequirePackage{color}

\begin{document}

\title{Numerical simulation of the electron capture process in a magnetar interior}
\slugcomment{Not to appear in Nonlearned J., 45.}
\shorttitle{The electron capture rate onto  protons }
\shortauthors{Z. F. Gao et al.}

\author{Z. F. Gao\altaffilmark{1,2}}
\altaffiltext{1}{Urumqi Observatory, NAOC, 40-5 South Beijing Road,
Urumqi Xinjiang, 830011, China zhifu$_{-}$gao@uao.ac.cn}
\altaffiltext{2}{Graduate University of the Chinese
   Academy of Scienes, 19A Yuquan road, Beijing, 100049, China}
\author{ N. Wang \altaffilmark{1}}
\affil{ Urumqi Observatory, NAOC, 40-5 South Beijing Road, Urumqi
Xinjiang, 830011, China}
\author{ J. P. Yuan \altaffilmark{1}}
\affil{Urumqi Observatory, NAOC, 40-5 South Beijing Road, Urumqi
Xinjiang, 830011, China}
\author{ L. Jiang \altaffilmark{3}}
\affil{ XinJiang Education Institute, 333 Guang Ming Road, Urumqi
Xinjiang, 830043, China }
\author{ D. L. Song \altaffilmark{4}}
\affil{ The Information Engineering University, 63 Science Road,
ZhengZhou Henan, 450001, China }

\begin{abstract}

In a superhigh magnetic field, direct Urca reactions can proceed
for an arbitrary proton concentration.  Since only the electrons
with high energy $E$ ($E > Q$, $Q$ is the threshold energy of
inverse $\beta-$decay) at large Landau levels can be captured, we
introduce the Landau level effect coefficient $q$ and the
effective electron capture rate $\Gamma_{\rm eff}$.  By using
$\Gamma_{\rm eff}$, the values of $L_{\rm X}$ and $L_{\rm \nu}$ are calculated,
where and $L_{\rm \nu}$, $L_{\rm X}$ are the average neutrino
luminosity of Anomalous X-ray Pulsars (AXPs) and the average X-ray
luminosity of AXPs, respectively.  The complete process of
electron capture inside a magnetar is simulated numerically.
\end{abstract}


\section{Introduction}
 Anomalous X-ray Pulsars (AXPs) and soft gamma-ray repeaters (SGRs) are a small group of
peculiar neutron stars (NSs) that are currently believed to have
superhigh magnetic fields B$\sim 10^{14}- 10^{15}$ G, and are hence
identified as magnetar candidates \citep{dun92, kou98, kou99, pac92,
tho95, tho96}.  Based on the convective dynamo mechanism, some
works have focused on the origin of the internal magnetic fields
of magnetars \citep{tho95,tho96}.  Magnetars are considered to be NSs
and might reasonably be expected to have very small radii R $\sim
10^{6}$ cm, although measured masses of these stars are about 1.4$
M_{\odot}$, where $M_{\odot}$ denotes the solar mass $\sim$ 2 $\times
10^{33}$ g.  In a common magnetar, the typical internal temperature is
$\sim 10^{8}$ K as estimated using the standard cooling mechanism \citep
{bac65,yak01,pag06}and its intrinsic superhigh magnetic fields may be
produced by the induced magnetic moments of the ${}^3P_2$ Cooper pairs
in an anisotropic neutron superfluid at a moderate lower interior
temperatures ($\sim$ 2.38 $\times 10^{8}$ K)\citep{pen06,pen09}.
Observations show that some SGRs and AXPs have a thermal type X-ray
flux, with the magnitude of average X-ray luminosity $L_{X}$ $\sim$
1.0$\times 10^{34}$- 5$\times 10^{36}$ erg~ s$^{-1}$.  A NS can be
treated as a system of magnetic dipoles ($B = B_{p}$, where $B_{p}$
is the polar magnetic field strength) because of the existence of a
${}^3P_2$ neutron superfluid in its interior.  In the presence of a
background magnetic field, the dipoles tend to become aligned in the
same direction, which would give rise to a phase transition from
paramagnetism to ferromagnetism in the interior of a NS if the
temperature drops below a critical temperature. Superhigh magnetic
fields of magnetars may originate from this phase transition and the
maximum field strength is about (3.0 -4.0) $\times 10^{15}$ G\citep{pen07}.

  When $B \gg B_{\rm cr}$, direct Urca reactions occur rapidly for
an arbitrary proton concentration due to the fact that strong magnetic
fields can alter matter compositions and increase phase space for
protons, with a resulting increase in $Y_{\rm e}$, where $Y_{\rm e}$ is the
mean electron number per baryon and $B_{\rm cr}$ is the quantum critical
magnetic field \citep{lai91}.  After entering the neutrino cooling
epoch, the direct Urca reaction is the simplest neutrino emission
process \citep{gam41,pet92}.  In the interior of a magnetar, where
matter is assumed to be totally transparent to neutrinos and
antineutrinos, the simple decay of neutrons and successive electron
capture take place simultaneously, as required by charge neutrality.
In order to provide a detailed view of the actual evolutionary scenario
of a magnetar, we should take into consideration the effective electron
capture rate $\Gamma _{\rm eff}$ (effective number of electrons captured
by one proton per second) due to the existence of Landau levels of
charged particles.

   The incorrect notion that $E_{\rm F}(\rm e)$ decreases with increasing
$B$ in an intense field ($B \gg B_{\rm cr}$) has been universally adopted
for a long time.  This misunderstanding has arisen because the solution
of the non-relativistic electron cyclotron motion equation $\hbar\omega_{B}$
is wrongly (or unsuitably) applied to calculate the energy state density
in a relativistic degenerate electron gas.  In addition, in some textbooks
on statistical physics, the torus located between the $n$-th Landau level
and the $(n+1)$-th Landau level is ascribed to the $n$-th Landau level
when calculating the statistical weight in momentum space.  If so, energy
(or momentum) will change continuously in the direction perpendicular to
the magnetic field, which is contradictory to the quantization of energy
( see Gao et al. 2010 for further details).  To the contrary, our point of
view is that in the case of an intense magnetic field, the stronger the
magnetic field, the higher the Fermi energy of electrons released from the
magnetic field energy.  The possible interpretations of this viewpoint
are also given in (Gao et al. 2010).

   This paper is organized as follows: in $\S$ 2 the values of
$\Gamma$ and $\langle E_{\rm n}\rangle$ are calculated; with regard to
the relationships between the magnetic fields and Landau levels,
we put forward our points of view and hypotheses in $\S$ 3.1;
in $\S$ 3.2 we calculate the magnitude of $q $ and give an
appropriate initial temperature $T_{0}$; in $\S$ 3.3 a numerical
simulation of the whole process of electron capture inside a
magnetar is presented; discussions and conclusions are given in
Section $4$; and the formula for the electron Fermi energy is deduced briefly in Appendix.

\section{The calculations of $\Gamma$ and $\langle E_{n} \rangle$ }

As the core density increases, the high electron Fermi energy drives
electron capture by nuclei and free protons, which reduces $Y_{\rm e}$ and
decreases the contribution of the degenerate electrons to the total
pressure supporting the core against gravitation collapse.  In this
paper, we focus on non-relativistic, degenerate nuclear matter and
ultra-relativistic, degenerate electrons under $\beta$-equilibrium
implying the following relationship among chemical potentials (called
the Fermi energies $E_{\rm F}$)of the particles: $\mu_{\rm p}+ \mu_{\rm e}= \mu_{\rm n}$, where
the neutrino chemical potential is ignored.  We assume that at zero-
temperature the NS is $\beta$-stable, but at non-zero temperature
($kT \ll E_{\rm F}(\rm i),\rm  i= n, p, e$, $k= $1.38$\times10^{-16}$ erg~ K$^{-1}$
is the Boltzmann constant), reactions $e^{-}+ p \rightarrow n+ \nu_e$
and $n\rightarrow e^{-}+ p + \nu^{-} _{e}$ proceed near the Fermi
energies $E_{\rm F}(\rm i)$ of the participating particles.  In the case of 0.5
$\rho_{0}\leq \rho\leq $ 2$\rho_{0}$, electrons are relativistic, neutrons
and protons are non-relativistic, and the following expressions are hold
approximately: $p^{2}_{\rm F}(\rm n)/2m_{\rm n}$ =60$(\rho/ \rho_{0})^{\frac{2}{3}}$ MeV,
$p^{2}_{\rm F}(p)/2m_{\rm p}$=1.9$(\rho/\rho_{0})^{\frac{4}{3}}$ MeV, $(m_{\rm n}-m_{\rm p})c^{2}$
= 1.29 MeV, where $\rho_{0}=$2.8 $\times 10^{14}$ g~ cm$^{-3}$ is the standard
nuclear density (see Chapter 11 of \citep{sha83}.

  For the purpose of
convenient calculation, we set $\rho=\rho_{0}$ in our model, yielding $Q =
E_{rm F}(\rm n)- E_{\rm F}(\rm p)=(m_{\rm n}-m_{\rm p})c^{2}+(p^{2}_{\rm F}(\rm n)/2m_{\rm n}- p^{2}_{\rm F}(\rm p)/2m_{\rm p})$=
59.4 MeV.  However, $E_{\rm \nu}> 0$ and the minimum of $Q$ is not less than
the minimum Fermi kinetic energy of the outgoing neutrons $p^{2}_{\rm F}(\rm n)/2m_{\rm n}$
=60 MeV, otherwise the process of $e^{-}+ p \rightarrow n + \nu_e$ will not occur.
Solving Eq.(9) in Appendix gives the expression $E_{\rm F}(\rm e)$= 40$(B/B_{\rm cr})
^{\frac{1}{4}}$ MeV when $B \gg B_{\rm cr}$ \citep{pen07,pen09}, where we assign
$\rho= \rho_{0}$ and $Y_{\rm e}\sim$ (0.08- 0.11) (see Appendix).  In this paper,
the range of $B$ is assumed to be (0.22346- 3.0)$\times 10^{15}$ G corresponding
to $E_{\rm F}(\rm e)\sim$ (60 -114.85) MeV, where 0.22346 $\times 10^{15}$ G is the
minimum strength of a superhigh magnetic field denoted as $B_{\rm f}$.  When $B$
drops below $B_{\rm f}$, the direct Urca process is quenched everywhere in the
magnetar interior.  The electron capture rates can be calculated by the
following equation:
\begin{eqnarray}
&&\Gamma=\frac{2\pi}{\hbar}\frac{G_{\rm F}^{2}C_{rm V}^{2}
(1+ 3a^{2})}{(2\pi^{2}\hbar^{3}c^{3})^{2}}I\nonumber\\
&&I=\int_{60}^{E_{\rm F}(\rm e)}(E_{\rm e}^{2}-m^{2}_{\rm e}
c^{4})^{\frac{1}{2}}E_{\rm e}\nonumber\\
&&(E_{\rm e}- 60)^{2}\frac{1}{e^{\frac{E{\rm e}-E_{\rm F}(\rm e)}
{kT}}+ 1}\frac{1}{e^{\frac{Q-E_{\rm e}}{kT}}+ 1}dE_{\rm e},
\end{eqnarray}
where $m_{\rm e}c^{2}$=0.511 MeV, $T= 1.0 \times 10^{8}$ K, and other terms appearing in Eq.
($1$) have already been defined in Chapter 18 of \citep{sha83}.  For convenience,
we use the symbol $\Lambda$ to represent $\frac{2\pi}{\hbar}\frac{G_{\rm F}^{2}
C^{2}_{\rm V}(1+ 3a^{2})}{(2\pi^{2}\hbar^{3}c^{3})^{2}}\approx$ 0.018 (MeV)$^{-5}$
s$^{-1}$.  If we want to determine $L_{\rm X}$ and the average kinetic energy of
the outgoing neutrons $\langle E_{\rm n}\rangle$, the average kinetic energy of
the outgoing neutrinos $\langle E_{\rm \nu}\rangle$ must firstly be calculated.  The
expression for $\langle E_{\rm \nu}\rangle$ is of the form:
\begin{equation}
\langle E_{\rm \nu} \rangle= \int_{Q}^{E_{\rm F}(\rm e)}S(E_{\rm e}- Q)^{3}
E_{\rm e}(E_{\rm e}^{2}-m_{\rm e}^{2}c^{4})^{\frac{1}{2}}dE_{\rm e}/I,
\end{equation}
where $I=\int_{Q}^{E_{\rm F}(\rm e)}S(E_{\rm e}-Q)^{2}E_{\rm e}(E_{\rm e}^{2}-m_{\rm e}^{2}c^{4})^{\frac
{1}{2}}dE_{\rm e}$.  By employing energy conservation via $E_{\rm \nu}= E_{\rm e}- Q$, $\langle
E_{\rm n} \rangle$ can be calculated by the equation $\langle E_{\rm n}\rangle = E_{\rm F}(\rm e)-
\langle E_{\rm \nu}\rangle-$1.29 MeV.  The calculation results are shown below in
tabular form.

\begin{table}
\small
\caption{The calculated  values of $\Gamma, \langle E_{\rm \nu} \rangle$
  and $\langle E_{\rm n}\rangle$\label{tb1-1}.}
\begin{tabular}{@{}crrrr@{}}
\tableline
 B       & $E_{\rm F}(\rm e)$  &   $\Gamma$ &  $\langle E_{\rm \nu}\rangle
 $&\multicolumn{1}{c}{$\langle E_{\rm n}\rangle$\tablenotemark{a}}\\
 (G)       &(MeV)      &(s$^{-1}$)     & (MeV)       &(MeV)\\
\tableline
3.0$\times10^{15}$&114.850  &1.022$\times10^{7}$& 43.20& 70.36 \\
2.8$\times10^{15}$&112.886 &8.892$\times10^{6}$ & 41.66& 69.94 \\
2.5 $\times10^{15}$&109.733 &7.043$\times10^{6}$ & 39.10 & 69.34 \\
2.0$\times10^{15}$&103.779 &4.367$\times10^{6}$& 34.31& 68.18 \\
1.5$\times10^{15}$&96.577 &2.255$\times10^{6}$& 28.54 & 66.75 \\
1.0 $\times10^{15}$&87.267&7.892$\times10^{5}$& 21.12 & 64.86\\
6.0 $\times10^{14}$&76.805  &1.501$\times10^{5}$& 12.88 & 62.64\\
3.7 $\times10^{14}$&68.805&13696  & 6.13&60.66\\
 3.5$\times10^{14}$&67.123& 9244.8 & 5.40& 60.43\\
 2.5$\times10^{14}$  &61.707&111.93 & 0.40 & $\sim$60\\
 2.25$\times10^{14}$&60.103&0.02375& 0.075& $\sim$60\\
\tableline
  \end{tabular}
\tablenotetext{a}{$\langle E_{\rm n} \rangle$ is calculated by
using the relation of\\
$\langle E_{\rm n}\rangle$=$E_{\rm F}(\rm e)$-$\langle E_{\rm \nu}\rangle$-1.29 MeV}
\end{table}

It can be seen from Table 1 that $\Gamma$ decreases with the decreasing
$E_{\rm F}(\rm e)$, and is reduced to zero when $E_{\rm F}(\rm e)= Q$.  The rate of the
electron capture reaction varies clearly with magnetic field strength:
when $B \gg B_{\rm f}$, $E_{\rm F}(\rm e)\gg Q$= 60 MeV, the reaction will happen
very quickly and the ${}^3P_2$ Cooper pairs will be destroyed immediately
by the outgoing neutrons in this process, which causes anisotropy in the
neutron superfluid and causes the induced magnetic field to disappear; when
the magnetic field weakens, the reaction rate becomes smaller; the reaction
will end if the field decreases below $B_{\rm cr}$.  By colliding with the
neutrons produced in the process $n+(n\uparrow n\downarrow)\longrightarrow
n+ n+ n$, the kinetic energy of the outgoing neutrons will be transformed
into thermal energy and then transformed into radiation energy via soft
X-ray and $\gamma$-ray emission \citep{pen07,pen09}.

   However, the values of $\Gamma$ in Table 1 are too large to represent
the full extent of electron capture in the interior of a magnetar.  We
believe that the calculated values of  $\Gamma$ from Eq.(1) are not the
actual values of $\Gamma$ and need to be modified.  The same should be
true of the neutron decay rate ($1/927$ s, calculated by Eq.(11.4.1) of
\citep{sha83}.  To explain this phenomenon, careful analysis found that
when Eq.(1)is used to calculate $\Gamma$, we simply assume that electrons
and protons are free when magnetic field effects are too weak to be
taken into consideration.  In other words, only when $B=0$ can Eq.(1) be
used, so the values of $\Gamma$ in Table 1 are simply the values of the
electron capture rate onto free protons.  When Eq.(11.4.1) of \citep{sha83}
is used to calculate neutron decay rate, we make a simple assumption: pure
neutron decay in a vacuum and strong interaction effects combined with the
magnetic field effects are ignored. Suppose the values of $\Gamma$ in Table 1
were not modified.  If reaction $e^{-}+ p \rightarrow n+ \nu_e$ proceeded at
the minimum rate of $\Gamma$ in Table 1, the value of $L_{\rm X}$ calculated by
Eq.(13) in $\S$ 3 would be far larger than the observed value ($\sim$1.0
$\times 10^{34}$- 5$\times 10^{36}$ erg~s$^{-1}$, see $\S$ 1.  Further,
electrons and protons would be depleted within a few minutes, which is not
consistent with the actual circumstances in a neutron star interior.  Instead,
$Y_{\rm e}$ decreases insignificantly in the whole process of electron capture,
and the mean value of $Y_{\rm e}$ of a neutron star is $\sim$ 0.05 and is
comparatively stable, which implies that the electron capture reaction
proceeds very slowly due to the existence of Landau levels.  In accordance
with the Pauli exclusion principle, electrons are situated in disparate
Landau levels from the lowest energy state (the ground energy state) up to
the highest energy state (the Fermi energy state), individually with the
overwhelming majority occupying $n= 0, 1$ Landau levels.  Only the electrons
occupying large Landau levels with high energy ($E > Q$) are allowed to
participate in the direct Urca process \citep{yak01}.  In other words,
whether an electron could be captured depends not only on the electron's
energy $E_{\rm e}$ but also on the number of the Landau level it occupies.  In
Eq.(1), $E_{\rm e}$ must be treated as a continuous function, otherwise $\Gamma$
cannot be calculated by using discrete Landau levels.  If $E_{\rm e}$ is treated
as continuous, as in the free-field case, we must modify $\Gamma$ by
introducing $q$ considering that the symmetry is broken in the momentum space
caused by the superhigh magnetic fields.  In the interior of an NS, different
forms of magnetic field (weakly quantizing field, strongly quantizing field
and non-quantizing field) could exist simultaneously.  The properties and
distributions of these fields, are described in more detail in (Gao et al.
2010).  Here, we focus on the magnetar interior, where the weakly quantizing
strong magnetic fields permit direct Urca processes.  The details will be given
in $\S$ 3.

\section{Numerical simulating the direct Urca process
in magnetar interior}

This section is composed of three subsections.  For each
subsection we present different methods and considerations.

\subsection{Our points of view and hypotheses}
For extremely strong magnetic fields, the cyclotron energy of an electron
becomes comparable to its rest-mass energy, and the transverse motion electron
becomes relativistic.  We can define a critical magnetic field (often called the
relativistic magnetic field) $B_{\rm cr}$ by the relation $\hbar\omega = m_{\rm e}c^{2}$
which gives $B_{\rm cr} = m^{2}_{\rm e}c^{3}/e\hbar$= 4.414$\times 10^{13}$ G.  In the
case of $B\geq B_{\rm cr}$, solving the relativistic Dirac Equation for electrons
gives the electron energy levels \citep{can77}:
\begin{equation}
E=[m^{2}_{\rm e}c^{4}(1 + \nu \frac{2B}{B_{\rm cr}})+p^{2}_{z}c^{2}]^{\frac{1}{2}},
\end{equation}
where $B$ is directed along the $z$-axis, quantum number $\nu$ is given by $\nu =
 n+ \frac{1}{2}+ \sigma$, $n = 0, 1, 2, \cdots $ is the Landau level number, and
spin $\sigma = \pm \frac{1}{2}, p_{z}$ is the $z$-component of the electron momentum.
Combining $B_{\rm cr} = m^{2}_{\rm e}c^{3}/e\hbar$ with $\mu_{\rm e}= e\hbar /2m_{\rm e}c$, we obtain
\begin{equation}
 E_{\rm e}^{2}(p_{z}, B, n, \sigma)= m_{\rm e}^{2}c^{4}+ p_{z}
 ^{2}c^{2}+(2n + 1 + \sigma)2m_{\rm e}c^{2}\mu_{\rm e}B,
\end{equation}
with $\mu_{\rm e} \sim$ 0.927 $\times 10^{-20}$ erg~ G$^{-1}$ being the magnetic moment of
an electron.  From Eq.(4), for electrons in a given Landau level, $E_{\rm e}$ increases
with $p_{z}c$;  if the value of $p_{z}c$ is invariable, $E_{\rm e}$ increases with
increasing Landau level number $n$.  In terms of the relation between the magnetic
fields and Landau levels, our points of view and hypotheses are as follows:
\begin{enumerate}
\item   Firstly, for electrons (or protons) in an intense magnetic field, the
maximum Landau level number, $n_{\rm m}$, decreases with increasing magnetic field strength
$B$.  However, it is very difficult to calculate exactly $n_{\rm m}$ occupied by a
homogeneous gas of cold electrons (or protons) in a given field, and so we can
only estimate it. Considering this limitation, we define a quantity $q_{\rm e}$, the ratio of the electron number in higher Landau levels to that in all Landau levels, written as
\begin{equation}
q_{\rm e}= \frac{1}{N_{tot}}(N(n)+N(n+1)+\cdots),
\end{equation}
where $n$ denotes the number of the lowest Landau level (not the ground Landau
level), below which electrons cannot be captured.
\item    Secondly, in the core of a NS, charge neutrality gives $n_{\rm e}= n_{\rm p}$.
Therefore, whenever the magnetic field significantly affects the electrons, it
also affects the protons, thus the values of $n_{\rm m}$ for electrons and protons
are essentially the same in a given magnetic field \citep{lai91}.  Similarly, we
define the quantity $q_{\rm p}$ as the ratio of the proton number in higher Landau
levels to that in all Landau levels, whose expression is the same as that of
$q_{\rm e}$.  We firstly introduce the Landau level effect coefficient $q =q_{\rm e}q_{\rm p}$,
and the effective electron capture rate $\Gamma_{\rm eff}$ is then defined as:
\begin{equation}
 \Gamma_{\rm eff} = q \Gamma = q_{\rm e}q_{\rm p}.
\end{equation}
\item   Finally, in the vicinity of the Fermi surface, the electrons with the
same energy $E$ could come from different Landau levels because the electrons
are degenerate.  However, the electrons occupying lower Landau levels cannot be
captured even if their energies are higher than the threshold reaction energy;
for higher Landau levels, there still exist some electrons with lower energies
$E$ ($E < Q$) that are not captured.

\end{enumerate}

In order to determine the order of magnitude of the Landau level effect
coefficient $q $ and to estimate the maximum Landau level number $n_{m}$ as
accurately as possible, we should firstly rewrite Eq.(4), as demonstrated in
$\S$ 3.2. we must modify $\Gamma$ by introducing $q$ considering that the symmetry 
is broken in the momentum space caused by the superhigh magnetic fields.

\subsection{The evaluations of $q$ and $T_{0}$}

When $B\gg B_{\rm cr}$, $p_{z}c\gg p_{\perp}c$ and $p_{z}c \gg m_{\rm e}c^{2}$
= 0.511 MeV, Eq.(4) can be well approximated as
\begin{equation}
E(\rm n)\approx p_{z}c\cdot(1+ \frac{1}{2}(\frac{m_{\rm e}
c^{2}}{p_{z}c})^{2}+(2n+ 1 +\sigma)\frac{m_{\rm e}c^{2}}
{p_{z}c}\frac{\mu_{\rm e}B}{p_{z}c}).
\end{equation}
Degenerate electron gas is distributed exponentially (a Maxwell
distribution)\citep{pen07}, so we define
\begin{eqnarray}
&&q_{n}(\rm e)=\frac{N(n)}{N(0)}=\exp\{-\frac{E(n)-E(0)}{kT}\}\nonumber\\
 &&=\exp\{-\frac{2n m_{\rm e}c^{2}\mu_{\rm e}B}{p_{z}c\cdot kT}\},
\end{eqnarray}
where $N(n)$ denotes the number of electrons in the $n$-th Landau level.
When $B\sim 10^{15}$ G, $T\sim 10^{8}$ K, $2m_{\rm e}c^{2}\mu_{\rm e}B \sim 10^{-5}$,
$p_{\rm F}(z)c\cdot kT \sim 10^{-6}$, $2m_{\rm e} c^{2}\mu_{\rm e}B/(p_{\rm F}(z)c\cdot kT)
\sim$ 10.  From these evaluations, it is clear that $N(n)\gg N(n+1)\gg N(n+2
\cdots)$, so $q_{n+2}(\rm e)\ll q_ {n+1}(\rm e)\ll q_{n}(\rm e)\ll$ 1.  According to the
Pauli exclusion principle, $N_{\rm tot}= n_{\rm e}$, here $N_{\rm tot}$ and$n_{\rm e}$ are
electron state density, electron number density, respectively.  Suppose $N(0)
\sim$ 0.9$n_{\rm e}$ in the ground state Landau level, then $N(0)/N_{\rm tot}\sim$
0.9.  Since $n_{\rm e}\sim 10^{36}$ cm$^{-3}$ \citep{sha83} and $N(n) \geq$ 1,
$n_{m}$ is estimated to be $\sim$ several or $\sim$ 10.  It is important to
note that, in a non-relativistic weak field, the electron cyclotron energy
is $\hbar\omega_{B}= \hbar eB/(m_{\rm e}c)$ =11.5 $B_{12}$ KeV, the maximum
Landau level number $n_{\rm m}\sim E_{\rm F}(\rm e)/\hbar\omega_{B}\sim 10^{2}$ or higher,
where $B_{12}$ is magnetic field in units of $10^{12}$ G; also, in the case
of a weakly quantizing relativistic strong magnetic field ($B\sim 10^{14}
\sim 10^{15}$ G), the solution of non-relativistic electron cyclotron motion
equation $\hbar\omega_{B}$ is no longer suitable, but if this equation is used,
the rest mass of an electron $m_{\rm e}$ must be replaced by its effective mass
$m^{*}_{\rm e}$, which is far larger than the former after taking into account the
effect of relativity. In this latter case $n_{m}$ could be estimated to be
$\sim$10 or higher, rather than 0 or 1, which shows our evaluations are
reasonable.  From the above discussion, we obtain the following approximate
relationship:
\begin{equation}
 q(\rm e)\approx \frac{1}{N_{\rm tot}} N(0)q_{n}(\rm e)=0.9 q_{n}(\rm e).
\end{equation}
Thus the electron number in the lowest Landau level can be accurately
approximated by
\begin{equation}
  N_{n}\approx 0.9 n_{\rm e}\exp\{-\frac{2n m_{\rm e}
 c^{2}\mu_{\rm e}B}{p_{\rm F}(z)c\cdot kT}\}.
\end{equation}
If we want to determine $q_{n}$, the value of $p_{\rm F}(z)c$ should be calculated
first.  Inserting $p_{z}c= p_{\rm F}(z)c$ and $E(\rm e)=E_{\rm F}(\rm e)$ into Eq.(4) gives
\begin{equation}
 [40(\frac{B}{B_{\rm cr}})^{\frac{1}{4}})]^{2} \approx m_{\rm e}^{2}
 c^{4}+ p^{2}_{\rm F}(z)c^{2}+(2n+ 1 + \sigma)2m_{\rm e}c^{2}\mu_{\rm e}B,
\end{equation}
with $p_{\rm F}(z)$ the highest momentum along the magnetic field.  Next, a way
of calculating the value of any $p_{\rm F}(z)c$ in Eq.(6) is introduced as follows.
For example, in the case of $B$ =3.0$\times 10^{15}$ G and $n$= 5, firstly, we
calculate the values of $p_{\rm F}(z)c$ corresponding to $\sigma$= 1 and $\sigma$
= -1, respectively, by using Eq.(6), then calculate the mean value of $p_{\rm F}
(z)c\sim$ 113.96 MeV.  From the analysis above, the effective electron capture
$\Gamma_{\rm eff}$ can be expressed as:
\begin{equation}
 \Gamma_{\rm eff}= q\Gamma= q(\rm e)q(\rm p)\Gamma=[0.9\exp\{-\frac{2n m_{\rm e}c^{2}
 \mu_{\rm e}B}{p_{\rm F}(z)c\cdot kT}\}]^2\Gamma .
\end{equation}
 
   The steady X-ray emission ($10^{34}\sim 10^{36}$ erg~ s$^{-1}$) could
be from the magnetar interior \citep{pen09}.  We assume that all protons take
part in the process of electron capture, and that all the kinetic energy of the
outgoing  neutrons is converted and radiated in the form of thermal energy, then
$L_{\rm X}$ can be expressed as
\begin{equation}
L=\Gamma_{\rm eff} n_{\rm p} V({}^3P_2) \langle E_{\rm n} \rangle,
\end{equation}
where $\langle E_{\rm n}\rangle\geq$ 60 MeV, otherwise the reaction ceases; $V({}
^3P_2)$ denotes the volume of ${}^3P_2$ anisotropic neutron superfluid, $V({}
^3P_2)=\frac{4}{3}\pi R_{5}^{3}$ cm$^{3}$, $R_{5}= 10^{5}$ cm, $\pi$ =3.14159,
and $n_{\rm p}= n_{\rm e}$= 9.6$\times 10^{35}$ cm$^{-3}$ setting $\rho= \rho_{0}$.  We
also gain an approximate expression of $T$ from Eqs.(12-13)
\begin{equation}
 T= \frac{2n m_{\rm e}c^{2}\mu_{\rm e}B}{p_{\rm F}(z)c\cdot k\ln[0.9(\frac
{\Gamma}{L_{\rm X}} n_{\rm e}\frac{4}{3}\pi R_{5}^{3}\langle E_{\rm n}
 \rangle)^{1/2}]}
\end{equation}

    In the initial stage of electron capture process, the initial X-ray
luminosity $L_{\rm X0}$ could be higher than the generally observed values
1$\times (10^{34}\sim$ 5$\times 10^{36}$ erg~ s$^{-1}$), so it is reasonable
to assume here that $L_{\rm X0}$ = 9$\times 10^{36}$ erg~ s$^{-1}$ and $B_{0}$=
3.0$\times 10^{15}$ G.  In order to estimate the order of magnitude of $q$
inside a neutron star, an appropriate initial temperature $T_{0}$ needs to
be determined.  Solving Eq.(10) gives the values of the possible initial
temperatures corresponding to $ n= 1, 2, 3, 4, 5, \cdots$, respectively.  The
calculation results are listed in Table 2.
\begin{table*}[t]
\caption{The possible values of the initial temperature}
\begin{tabular}{@{}lllll}
\hline
$\rm  n$ & $p_{F}(z)c$&\multicolumn{1}{c}{$T$\tablenotemark{a}}
&\multicolumn{1}{c}{$q$\tablenotemark{b}}&\multicolumn{1}{c}
{$\Gamma_{\rm eff}$\tablenotemark{c}} \\
&(MeV) &(K) & ($\times 10^{-18}$)& ($\times 10^{-11}$s$^{-1})$\\
 $\rm  1$ & $114.62$& 8.85795$\times10^{7}$ &1.9543& 1.9881\\
$\rm  2$ & $114.46$&1.77407$\times10^{8}$  & 1.9543 & 1.9881\\
$\rm  3$ & $114.31$& 2.66459$\times10^{8}$  & 1.9543 & 1.9881\\
$\rm  4$ & $114.15$&3.55777$\times10^{8}$  & 1.9543  & 1.9881\\
$\rm  5$ & $113.99$&4.45345$\times10^{8}$ & 1.9543 & 1.9881\\
$\rm  6$ & $113.84$&5.35118$\times10^{8}$   & 1.9543 & 1.9881\\
\hline
\end{tabular}
\tablenotetext{a}{$T= \frac{2n m_{\rm e}c^{2}\mu_{\rm e}B}{p_{\rm F}
(z)c\cdot k\ln[0.9(\frac{\Gamma}{L_{\rm X}}
 n_{\rm e} \frac{4}{3}\pi R_{5}^{3}\langle E_{\rm n} \rangle)^{1/2}]}$}
\tablenotetext{b}{$q =\frac{L_{\rm X}}{n_{p}\Gamma V({}^3P_2)
\langle E_{\rm n} \rangle}$}
\tablenotetext{c}{$\Gamma_{\rm eff}=\frac{L_{\rm X}}{n_{\rm p}
 V({}^3P_2) \langle E_{\rm n} \rangle}$,
$\langle E_{\rm n} \rangle \sim$ 70.36 MeV, assuming that \\
the initial X-ray luminosity $L_{\rm X0}$= 9$\times 10^{36}$
erg~s$^{-1}$ \\
and the initial magnetic field strength
$B_{0}$=3.0$\times 10^{15}$ G}
\end{table*}

    According to neutron star cooling theory, the typical magnetar
internal temperature is about 3$\times 10^{8}$ K \citep{yak01}.  When
$B \sim$ 1.0$\times 10^{15}$ G, $T\sim 10^{8}$ K, but $T < T_{C}({}
^3P_2)$, the critical temperature of ${}^3P_2$ a neutron superfluid,
$T < T_{C}({}^3P_2)= \Delta_{\rm max}({}^3P_2)/2k \approx$ 2.78$\times
10^{8}$ K, where $\Delta_{\rm max}({}^3P_2)\sim$ 0.05 MeV is the energy
gap maximum of ${}^3P_2$ \citep{pen06,pen09}.  From Table 2, an
appropriate initial temperature($\sim$ 2.6646$\times 10^{8}$ K) is
selected by considering that the calculated values for $n$ = 1, 2 are
too low , whereas the calculated values for $n$ =4, 5, 6  are too high
(higher than $T_{C}({}^3P_2)$).  We must stress that this initial
temperature is selected arbitrarily for a particular case and may be
different for other magnetars.   From Table 2, once $L_{\rm X0}$ has been
calculated, $q$ is determined ($\sim$ 1.9543$\times 10^{-18}$) but has
little effect on $n$ and $T$.  It is worth noting that the calculated
values of temperature are the possible values (or the expected values)
of the initial temperature, not the variable range of temperature.  In
the vicinity of the Fermi surface, the electrons with the same energy
could come from different Landau levels because the electrons are
degenerate.   However, not all the electrons near the Fermi surface
can be captured, since electrons with energy ($E_{\rm n}> Q$ =60 MeV)
cannot be captured if their Landau level number is too small ($n <3$);
similarly, some electrons occupying Landau levels ($n\geq$ 3)are not
captured because of their lower energy ($E < Q$).  We next define the
effective captured electron number of the $n$-th Landau level $N_{\rm eff}(n)$,
\begin{equation}
 N_{\rm eff}(n)= N(n)- N_{E= Q}(n),
\end{equation}
where $N_{E= Q}(n)$ denotes the number of electrons with energy $E$
($E_{n} < Q$) in the $n$-th Landau level.  Accordingly, the expression
for $q_{n}$ should be modified:
\begin{equation}
q_{n}=\exp\{-\frac{2n m_{\rm e}c^{2}\mu_{\rm e}B}{p_{\rm F}(z)c\cdot kT}\}
-\exp\{-\frac{2n m_{\rm e}c^{2}\mu_{\rm e}B}{p_{E= Q}(z)c\cdot kT}\}.
\end{equation}
If $B$ = 3.0$\times 10^{15}$ G, $T$ = 2.6646$\times 10^{8}$ K,
 $n $= 3, $E$ = 60 MeV,  from Eq.(7), we get $p_{E=Q}(z)c$
= 59.102 MeV, $\exp\{-\frac{2n m_{\rm e}c^{2}\mu_{\rm e}B}{p_{E=Q}(z)c
\cdot kT}\}$= 9.141$\times 10^{-18}$.  Further, when $E_{\rm F}(\rm e)$ =
114.85 MeV, then $p_{\rm F}(z)c$=114.31 MeV, $\exp\{-\frac{2n m_{\rm e}
c^{2}\mu_{\rm e}B}{p_{\rm F}(z)c\cdot kT}\}$=1.549 $\times 10^{-9}$.  In a
given Landau level, similar to a mushroom cloud, the electron
number increases exponentially with increasing $p(z)c$, so $q_{n}(\rm e)
\approx \exp\{-\frac{2n m_{\rm e}c^{2}\mu_{\rm e}B}{p_{\rm F}(z)c\cdot kT}\}$, which
illustrates our approximate ways above are reasonable.

\subsection{Numerical simulation of a complete electron capture process}
In this section, a simple way of simulating magnetar cooling and magnetic
field decay is introduced briefly.  For an  example, in order to determine the
temperature corresponding to $B$=2.8 $\times 10^{15}$ G, by combining $q_{4}$ =
$[0.9\exp\{-4 \frac{2m_{\rm e}c^{2}\mu_{\rm e}B}{p_{\rm F}(z)c\cdot kT}\}]^2$ with
$p_{\rm F}(z)c$= 112.37 MeV, $T$ is decreased by step $\Delta T$ =0.0001 $\times
10^{8}$ K from an initial temperature $T_{0}$.  The expected value of $T$ is reached when
the value of $q$ is just below 1.9543$\times 10^{-18}$.  Once $T$ is determined,
the relevant values of $L_{\rm X}$, $q$ and $\Gamma_{\rm eff}$ can be calculated
easily.  The details of numerical simulations are shown in Table 3.
\begin{table*}[t]
\caption{The details of numerical
simulating magnetar cooling}
\begin{tabular}{@{}llll}
\hline
$T$          & $q$   &  $\Gamma_{\rm eff}$  & $L_{\rm X}$ \\
($\times 10^{8}$ K)&($\times 10^{-18}$)& (s$^{-1}$)&(erg s$^{-1}$\\
\hline
 2.5303 & 1.9582& 1.7412$\times10^{-11}$  & 7.835$\times10^{36}$ \\
 2.5302& 1.9551 & 1.7385$\times10^{-11}$ & 7.821$\times10^{36}$  \\
2.5301\tablenotemark{a}&1.9519\tablenotemark{a}&1.7356$\times10^{-11}
$\tablenotemark{a}  & 7.810$\times10^{36}$\tablenotemark{a} \\
 2.5300& 1.9488 & 1.7330$\times10^{-11}$  & 7.798$\times10^{36}$\\
\hline
\end{tabular}
\tablenotetext{a}{ The possible values of related
quantities in physics}
\end{table*}
where $\langle E_{\rm n}\rangle$= 69.94 MeV, $n$ = 3, $\Gamma$=8.892
$\times 10^{6}$ s$^{-1}$.  By using the same method, the values of
$q$, $\Gamma_{\rm eff}$, $T$ and $L_{rm X}$ in different stages of the
electron capture process are calculated and listed.  The details of numerical simulation of the whole process of electron
capture inside magnetar are shown in Table 4.

\begin{table*}[t]
\small
\caption{Numerically simulating the whole process of electron
capture inside magnetar. \label{tb4-2}}
\begin{tabular}{@{}crrrrr@{}}
\tableline
$B$  &  $p_{\rm F}(z)c$ & $T$&\multicolumn{1}{c}{$q$\tablenotemark{a}}
 & $\Gamma_{eff}$ &\multicolumn{1}{c}{$L$\tablenotemark{b}}\\
(G)  & (MeV)& (K)& s$^{-1}$ &  s$^{-1}$ & erg~ s$^{-1}$\\
\tableline
3.0$\times10^{15}$ & 114.31 & 2.6646$\times10^{8}$ & 1.9543
$\times10^{-18}$& 1.4881$\times10^{-11}$&9.0$\times10^{36}$\\
2.8$\times10^{15}$ & 112.37 & 2.5301$\times10^{8}$ & 1.9519
$\times10^{-18}$&1.7356$\times10^{-12}$&7.810$\times10^{36}$\\
2.5$\times10^{15}$ & 109.26 & 2.3233$\times10^{8}$ & 1.9513
$\times10^{-18}$ &1.3743$\times10^{-11}$&6.131$\times10^{36}$\\
2.0$\times10^{15}$& 103.38 & 1.9643$\times10^{8}$ &1.9491
$\times10^{-18}$& 8.5117$\times10^{-12}$& 3.734$\times10^{36}$\\
1.5$\times10^{15}$ & 96.15  & 1.5840$\times10^{8}$ & 1.9489
$\times10^{-18}$& 4.3948$\times10^{-12}$&1.887$\times10^{36}$\\
1.0$\times10^{15}$ & 87.03  & 1.1666$\times10^{8}$ & 1.9448
$\times10^{-18}$&1.5348$\times10^{-12}$& 6.405$\times10^{35}$\\
6.0 $\times10^{14}$ & 76.64  & 7.9485$\times10^{7}$ & 1.9446
$\times10^{-18}$ & 2.9188$\times10^{-13}$& 1.176$\times10^{35}$\\
3.7$\times10^{14}$ & 67.95 & 5.5284$\times10^{7}$ & 1.9441
$\times10^{-18}$&2.6626$\times10^{-14}$&1.039$\times10^{34}$\\
 3.5$\times10^{14}$ & 66.01 & 5.3028$\times10^{7}$ & 1.9423
$\times10^{-18}$ & 1.7956$\times10^{-14}$&6.982$\times10^{33}$\\
2.5$\times10^{14}$ & 61.62 & 4.1190$\times10^{7}$ & 1.9417
$\times10^{-18}$ &2.1733$\times10^{-16}$&8.390$\times10^{31}$\\
 2.25$\times10^{14}$ &60.02 & 3.8059$\times10^{7}$ & 1.9412
$\times10^{-18}$ &4.6104$\times10^{-20}$&1.780$\times10^{28}$\\
\tableline
\end{tabular}
\tablenotetext{a}{$q$= $[0.9\exp\{-\frac{2n m_{\rm e}c^{2}
\mu_{\rm e}B}{p_{\rm F}(z)c\cdot kT}\}]^2 $}
\tablenotetext{b}{$L_{\rm X}= \Gamma_{\rm eff} n_{\rm p} V({}^3P_2)
\langle E_{\rm n} \rangle$}
\end{table*}

From the simulations above, we infer that, once the value of
$L_{\rm X0}$ is given, $q$ decreases insignificantly and can be
treated as a constant which could be explained as follows: on the
one hand, in a magnetar interior, the electrons are extremely
relativistic and degenerate ($E_{\rm e}\gg m_{\rm e}c^{2}, E_{\rm F}(\rm e)\gg
kT$), such that when $T$ falls, electron transition between Landau
levels is not permitted because the electrons can be treated as
having a zero-temperature approximation; on the other hand, the
processes of electron capture and $\beta$-decay occur at the same
time (as required by electrical neutrality), so when $B$ decays,
the depleted protons and electrons are recruited for many times
leading to only a small decrease in the value of $Y_{\rm e}$ .  In
order to validate our speculations further, we can assume
$L_{\rm X0}$=1.0$\times10^{36}$ erg~ s$^{-1}$, and simulate the whole
process of  electron capture numerically in the same way.  The
details are shown in Table 5.

\begin{table*}[t]
\small
\caption{Numerical simulating the whole process of electron
capture inside magnetar.\label{tb5-2}}
\begin{tabular}{@{}crrrrr@{}}
\tableline
$B$& $p_{\rm F}(z)c$&$T$ &\multicolumn{1}{c}{$q$\tablenotemark{a}}
& $\Gamma_{\rm eff}$ &\multicolumn{1}{c}{$L$\tablenotemark{b}} \\
(G) & (MeV)& (K)  & s$^{-1}$ & s$^{-1}$  & erg~ s$^{-1}$\\
\tableline
3.0$\times10^{15}$ & 114.31 & 2.5277$\times10^{8}$ & 2.1614
$\times10^{-19}$& 2.2090$\times10^{-12}$& 1.0$\times10^{36}$\\
2.8$\times10^{15}$& 112.37 & 2.3999$\times10^{8}$& 2.1610
$\times10^{-19}$&1.9516$\times10^{-13}$& 8.647$\times10^{35}$\\
2.5$\times10^{15}$& 109.26 & 2.2037$\times10^{8}$&2.1585
$\times10^{-19}$&1.5202$\times10^{-13}$&6.782$\times10^{35}$\\
2.0$\times10^{15}$& 103.38 &1.8632$\times10^{8}$ & 2.1568
$\times10^{-19}$&9.4188$\times10^{-13}$&4.132$\times10^{35}$\\
1.5$\times10^{15}$& 96.15 & 1.50247$\times10^{8}$ & 2.1563
$\times10^{-19}$&4.8625$\times10^{-13}$&2.088$\times10^{35}$\\
1.0$\times10^{15}$& 87.03 & 1.1066$\times10^{8}$& 2.1555
$\times10^{-19}$ & 1.7011$\times10^{-13}$& 7.098$\times10^{34}$\\
6.0 $\times10^{14}$& 76.64 & 7.5397$\times10^{7}$ & 2.1552
$\times10^{-19}$& 2.3250$\times10^{-14}$& 1.304$\times10^{34}$\\
 3.7$\times10^{14}$ & 67.95 & 5.2440$\times10^{7}$ & 2.1535
$\times10^{-19}$&2.9494$\times10^{-15}$ & 1.151$\times10^{33}$\\
 3.5$\times10^{14}$ & 66.01 & 5.0301$\times10^{7}$ & 2.1531
$\times10^{-19}$& 1.9905$\times10^{-15}$& 7.739$\times10^{32}$\\
 2.5$\times10^{14}$ & 61.62 & 3.9072$\times10^{7}$ & 2.1529
$\times10^{-19}$&2.4097$\times10^{-17}$&9.303$\times10^{30}$\\
2.25$\times10^{14}$& 60.02 & 3.6102$\times10^{7}$ & 2.1523
$\times10^{-19}$& 5.1117$\times10^{-21}$ &1.974$\times10^{27}$\\
\tableline
\end{tabular}
\tablenotetext{a}{$q$= $[0.9\exp\{-\frac{2n m_{\rm e}c^{2}
\mu_{\rm e}B}{p_{\rm F}(z)c\cdot kT}\}]^2 $}
\tablenotetext{b}{$L= \Gamma_{\rm eff} n_{\rm p} V({}^3P_2)
\langle E_{\rm n}\rangle$}
\end{table*}

In reality, the observed values of $L_{\rm X}$ are different for
different magnetars.  However, for most magnetars, $B\sim 10^{14}
\sim 10^{15}$ G, $T\sim 10^{7}-10^{8}$ K, so the observed X-ray
luminosity $ \sim10^{34 }\sim 10^{36}$ erg~s$^{-1}$ which can be
explained by the calculations above.  The values of the average
neutrino luminosity of AXPs, $L_{\rm \nu}$, are determined as follows:
\begin{equation}
 L_{\rm \nu}=\Gamma_{\rm eff} n_{\rm p} V({}^3P_2) \langle E_{\rm \nu} \rangle.
\end{equation}
Clearly, our approach to calculating $L_{\rm \nu}$ is completely different
from previous methods.  For instance, the neutrino emissivity $Q_{\nu}$
can be calculated by Eq.(14) of \citep{bai99} as follows:
\begin{eqnarray}
&&Q= Q^{0}_{\rm \nu}\times R^{qc}_{B}\nonumber\\
&&Q^{0}_{\rm \nu}=\frac{457\pi G^{2}(1+ 3g^{2}_{A})}
{10080}m^{\ast}_{n}m^{\ast}_{p}\mu_{\rm e}T_{9}^{6},
\end{eqnarray}
where $T_{9}= T/10^{9}$ K, $m^{\ast}_{n}$ and $m^{\ast}_{p}$ are
nucleon effective masses in dense matter, $G=G_{F}\cos\theta_{C}$,
$G_{F}$=1.436$\times10^{-49}$ erg~ cm$^{3}$ is the Fermi weak coupling
constant, $\theta_{C}\approx 13^{o}$ is the Cabibbo angle, $g_{A}$=
1.261 is the axial-vector coupling constant, $Q^{0}_{\nu}$ is the
field-free emissivity, the factor $R^{qc}_{B}$ describes the effect
of the magnetic field.  In Eq.(18), $E_{\rm \nu}$ is $\sim kT_{9}$ and
$B$ can be as high as $B\geq 10^{18}$ G, but is not discussed (here
the range of $B$ is 0$\sim$ 3$\times 10^{16}$ G).  Thus the value of
$E_{\rm \nu}$($E_{\rm \nu}\sim$ 0.086 MeV, when $T\sim 10^{9}$ K) is far
less than those of Table 1 in $\S$ 2, and the range of $B$ is not
consistent with that of our model.  We therefore ask what is the
essential difference between $E_{\nu}$ of the direct Urca reaction
and that of the modified Urca reaction (in the process of a modified
Urca reaction, $E_{\rm \nu}\sim kT_{9}$).  It is universally acknowledged
that the neutrino emissivity of the modified Urca reaction is about
six orders of magnitude smaller than that of the direct Urca reaction.
May this  be because of the increase of the number of particles
participating in the direct Urca reaction while the average kinetic
energy of the outgoing neutrinos $\langle E_{\rm \nu}\rangle$ is invariable
in both cases?  Ultrastrong magnetic fields can generate a noticeable
magnetic broadening of the direct Urca process threshold and the
thresholds of other reactions in a magnetar interior \citep{yak01}.
However, the modified Urca process are negligible if the direct Urca
process is allowed.  What is more, it is easy to imagine that the
neutrino flux comes from the thermal energy in the magnetar interior,
rather than from the free energy of the superhigh magnetic field, if
$E_{\rm \nu}\sim kT_{9}$ is used in the direct Urca reaction.  There
exist many different cooling mechanisms, including neutrino emission,
inside a NS.  In the direct Urca reaction, $\langle E_{\rm \nu}\rangle$
and $L_{\rm \nu}$ are ultimately determined by $B$, and are weak functions
of $T$, which is only equivalent to background temperature and decreases
with decreasing $B$.  The $\beta-$decay and related reactions in strong
magnetic fields have been investigated since the late 1960s (e.g.,
\citep{can71,deb98,lai91} and references therein).  Despite this previous
research, existing analysis is approximate and many assumptions are
invoked consequently, because our computational methods are completely
new approaches, that need to be validated empirically.  By using Eq.(11),
schematic diagrams of the neutrino luminosity $L_{\rm \nu}$ as a function
of magnetic field strength $B$ are shown in Figure 1

\begin{figure}
 \includegraphics[width=0.75\columnwidth,angle=-90]{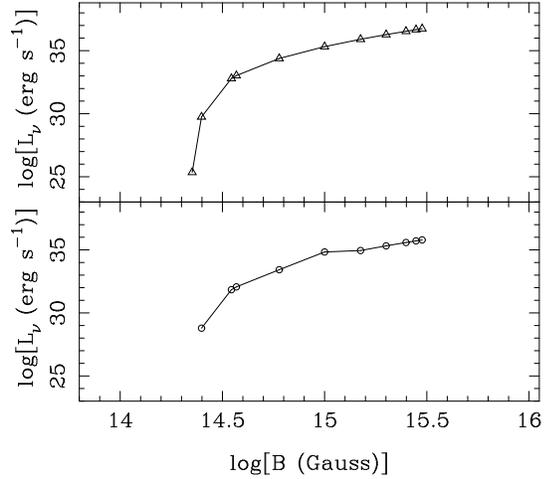}
 \caption{The schematic diagrams of the neutrino luminosity $L_{\rm \nu}$
as a function of magnetic field strength $B$.   Circle and triangle
 mark the values of variables corresponding to $L_{\rm X0}$= 9.0$\times 10^{36}$
erg~ s$^{-1}$, $L_{\rm X0}$= 1.0$\times 10^{36}$ erg~ s$^{-1}$, respectively.}
  \label{fig:bl}
 \end{figure}

From Table 1 and Tables 4-5, we find that the neutrino luminosity
$L_{\rm \nu}$ is also a weak function of the background temperature
$T$.  The schematic diagrams of the neutrino luminosity $L_{\rm \nu}$
versus the background temperature $T$ are plotted in Figure 2.

\begin{figure}
 \includegraphics[width=0.75\columnwidth,angle=-90]{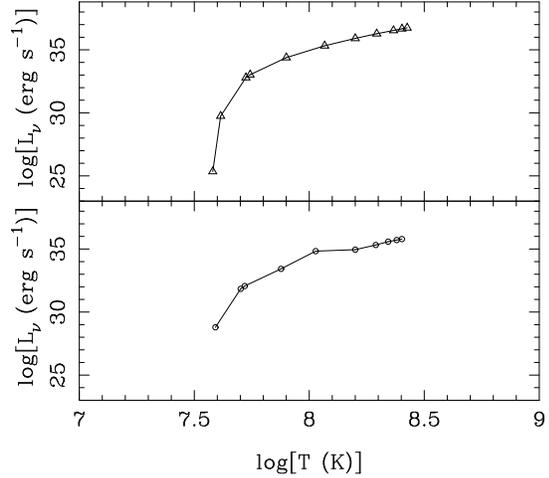}
 \caption{The schematic diagrams of the neutrino luminosity $L_{\rm \nu}$
as a weak function of temperature $T$.   Circle and triangle
 mark the values of variables corresponding to $L_{X0}$= 9.0$\times 10^{36}$
erg~ s$^{-1}$, $L_{\rm X0}$= 1.0$\times 10^{36}$ erg~ s$^{-1}$, respectively.}
 \label{fig:tl}
 \end{figure}

Figure 2 illustrates that the neutrino emission decreases with
falling temperature. The relationship between magnetic field
strength $B$ and the background temperature $T$ is shown in Figure
3, based on the data in Tables 4-5.

 \begin{figure}[th]
 \includegraphics[width=0.75\columnwidth,angle=-90]{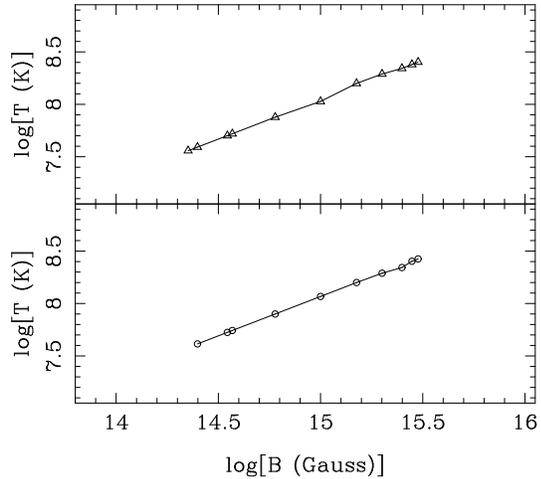}
\caption{The schematic diagrams of temperature $T$
as a weak function of magnetic field strength $B$.   Circle and triangle
 mark the values of variables corresponding to $L_{\rm X0}$= 9.0$\times 10^{36}$
erg~ s$^{-1}$, $L_{\rm X0}$= 1.0$\times 10^{36}$ erg~ s$^{-1}$, respectively.}
\label{fig:bt}
 \end{figure}

Figure 3 shows that the direct Urca reaction proceeds as long as
the magnetic field $B$ is high than the critical value $B_{\rm cr}$
corresponding to the minimum value of the background temperature
$T$ in the core, and when $T$ falls further with decreasing $B$ to
below the critical value, the neutrino luminosity and direct Urca
process are quenched everywhere in the core.  It should be noted
that the surface temperature is controlled by crustal physics, and
is independent of the evolution of the core.  In these three
figures, $B$ is assumed to be in the range 2.24$\times 10^{14}$ G
-3.0$\times 10^{15}$ G. Note that when $B\leq B_{\rm f}$, the direct
Urca processes cease, while the modified Urca processes still
occur, producing the weaker X-ray and neutrino flux.

\section{ Discussions and Conclusions}

In this paper, we introduce an approximate method for
investigating the effects of Landau levels on the evolution of a
superhigh magnetic field, and also numerically simulate the
process of magnetar cooling and magnetic field decay.  The main
conclusions are as follows:

1. The effect of a superhigh magnetic field would be to speed up
the cooling of magnetar. When $B$ decays, the values of $\Gamma$,
$T$ and $Y_{\rm e}$ decrease, but$\Delta Y_{\rm e}$ could be very small.

 2. In the magnetar interior, the ${}^3P_2$ Cooper pairs will be destroyed
 quickly by the outgoing neutrons via the process of electron capture,
 so the induced magnetic field will disappear.

 3. The abnormal X-ray flux $L_{\rm X}$ and neutrino flux $L_{\nu}$
 come from the free energy of the superhigh magnetic field, not from
 the thermal energy in the core of magnetar, and are all ultimately
 determined by the magnetic field strength.

Finally, we are hopeful that our assumptions and numerical
simulations can be combined with observations in the future, to
provide a deeper understanding of the nature of the superhigh
magnetic fields of magnetars.

\acknowledgments
The author Z. F. Gao is very grateful to Prof.Qiu-He. Peng in
Department of Astronomy, Nanjing University for his help of
improving the paper.  This work is supported by Xinjiang Natural
Science Foundation No. 2009211B35, the key Directional Project of
CAS and NNSFC under the project No. 10173020, No. 0673021 and                                                                                                              Chinese National Science Foundation through grant No. 10573005.

\appendix
\textbf{Appendix}
\section{The calculation of  $ E_{F }(e)$ in the presence
of magnetic field}

By summing over electron energy states (per unit volume)in a 6-dimension
 phase space, we can express $N_{\rm pha}$ as follows
 \begin{eqnarray}
   N_{\rm pha}=&& \frac{2\pi}{h^{3}}\int dp_{z}\sum_{n=0}^{n_{\rm m}(p_z,
      \sigma,B^{*})}\sum g_{n}\nonumber\\
       &&\int \delta(\frac{p_{\perp}}{m_{\rm e}c}-[(2n+ 1+ \sigma)B^{*} ]
         ^{\frac{1}{2}}) p_{\perp}dp_{\perp}
      \end{eqnarray}
in which the Dirac $\delta$-function and the relation $2\mu_{\rm e}
B_{\rm cr}/m_{\rm e}c^{2}$= 1 are used, $B^{*} = B/B_{\rm cr}$ is a non-dimensional
magnetic field, $p_{\perp}= m_{\rm e}c\sqrt{(2n+ 1+ \sigma)B^{*}}$ denotes electron
momentum perpendicular to the magnetic field, $2\mu_{\rm e}B_{\rm cr}/m_{\rm e}c^{2}$
= 1.  For $n$ = 0, the spin is antiparallel to $B$, the spin quantum number
$\sigma = -1$, so the ground state Landau level is non-degenerate; whereas at
higher levels $n \geq$ 1 are doubly degenerate, and the spin quantum number
$\sigma = \pm$ 1.  Therefore the spin degeneracy $g_{n} = 1$ for $n$ = 0 and
$g_{n}$ = 2 for $n \geq$ 1, then Eq.(A1) can be rewritten

\begin{eqnarray}
&&N_{\rm pha}= 2\pi(\frac{m_{\rm e}c}{h})^{3}\int d(\frac{p_{z}}{m_{\rm e}c})
      [\sum_{n=0}^{n_{m}(p_z, \sigma, B^{*})}\nonumber\\
 &&\int \delta(\frac{p_{\perp}}{m_{\rm e}c}-(2n B^{*} )^{\frac{1}{2}})
 (\frac{p_{\perp}}{m_{\rm e}c})d(\frac{p_{\perp}}{m_{\rm e}c})\nonumber\\
 &&+\sum_{n=1}^{n_{m}(p_z, \sigma,B^{*})}\int \delta(\frac{p_{\perp}}
 {m_{\rm e}c}-(2(n+ 1)B^{*})^{\frac{1}{2}})(\frac{p_{\perp}}
 {m_{\rm e}c})d(\frac{p_{\perp}}{m_{\rm e}c})]
      \end{eqnarray}

   The maximum Landau level number $n_{m}$ is the upper limit of the summation
over $n$ in Eq.(A2), which is uniquely determined by the condition $(p_{\rm F}
(z)c)^{2}\geq$ 0 \citep{lai91}. The expression for $n_{m}$ is

      \begin{equation}
     n_{\rm m}(p_{z},B^{*},\sigma = -1)= Int[\frac{1}{2B^{*}}[(\frac
     {E_{\rm F}}{m_{\rm e}c^{2}})^{2} -1 -(\frac{p_{z}}{m_{\rm e}c})^{2} ]]
     \end{equation}
      \begin{equation}
     n_{\rm m}(p_{z},B^{*},\sigma = 1)=Int[\frac{1}{2B^{*}}[(\frac{E_{\rm F}}
     {m_{\rm e}c^{2}})^{2} -1 -(\frac{p_{z}}{m_{\rm e}c})^{2} ]- 1]
     \end{equation}

where $Int[x]$ denotes an integer value of the argument $x$.  After a
complicated process, Eq.(A6) may now be rewritten
\begin{eqnarray}
&&N_{\rm pha}=6\pi\sqrt{2B{*}}(\frac{m_{\rm e}c}{h})^{3}\int_{0}^{\frac{E_{\rm F}}
{m_{\rm e}c^{2}}}n^{\frac{3}{2}}_{m}(p_z,B^{*})\nonumber\\
&&d(\frac{p_{z}}{m_{\rm e}c})-2\pi(\frac{m_{\rm e}c}{h})^{3}\sqrt{2B{*}}(\frac{E_{\rm F}}{m_{\rm e}c^{2}})\nonumber\\
&&=6\pi\sqrt{2B{*}}(\frac{m_{\rm e}c}{h})^{3}(\frac{1}{2B^{*}})^{\frac{3}{2}}
\int_{0}^{\frac{E_{\rm F}}{m_{\rm e}c^{2}}}[(\frac{E_{\rm F}}{m_{\rm e}c^{2}})^{2} -1 -
(\frac{p_{z}}{m_{\rm e}c})^{2}]^{\frac{3}{2}}\nonumber\\
&&d(\frac{p_{z}}{m_{\rm e}c})-2\pi(\frac{E_{\rm F}}{m_{\rm e}c^{2}})(\frac{m_{\rm e}c}{h})^{3}\sqrt{2B{*}}\nonumber\\
&&=\frac{3\pi}{B^{*}}(\frac{m_{\rm e}c}{h})^{3}\int_{0}^{\frac{E_{\rm F}}{m_{\rm e}c^{2}}}
 [(\frac{E_{\rm F}}{m_{\rm e}c^{2}})^{2} -1-(\frac{p_{z}}{m_{\rm e}c})^{2}]^{\frac{3}{2}}\nonumber\\
 &&d(\frac{p_{z}}{m_{\rm e}c})-2\pi(\frac{E_{\rm F}}{m_{\rm e}c^{2}})(\frac{m_{\rm e}c}{h})^{3}\sqrt{2B{*}}
  \end{eqnarray}
In order to deduce the formula for $E_{\rm F}(\rm e)$, we firstly
introduce two non-dimensional variables $\chi$ and $\gamma_{\rm e}$,
which are defined as $\chi =(\frac{p_{z}}{m_{\rm e}c})/(\frac{E_{\rm F}}
{m_{\rm e}c^{2}})= p_{z}c/E_{\rm F}$ and $\gamma_{\rm e}= E_{\rm F}/m_{\rm e}c^{2}$,
respectively, then Eq.(A5) can be rewritten as

\begin{eqnarray}
&&N_{\rm pha}=\frac{3\pi}{B^{*}}(\frac{m_{\rm e}c}{h})^{3}(\gamma_{\rm e})^{4}
\int_{0}^{1}(1-\frac{1}{\gamma^{2}_{\rm e}}-\chi^{2})^{\frac{3}{2}}d\chi \nonumber\\
&&- 2\pi \gamma_{\rm e}(\frac{m_{\rm e}c}{h})^{3}\sqrt{2B{*}}
 \end{eqnarray}
The electron number density is determined by
\begin{equation}
n_{\rm e}  =   N_{A}\rho Y_{\rm e}
\end{equation}
where $N_{A}$= 6.02$\times 10^{23}$ is the Avogadro constant
\citep{sha83}.  For a given nucleus with proton number $Z$ and
nucleon number $A$, the relation $Y_{\rm e}= Z/A$ always holds.  Combining
Eq.(A5) with Eq.(A6), we find
\begin{eqnarray}
&&\frac{3\pi}{B^{*}}(\frac{m_{\rm e}c}{h})^{3}(\gamma_{\rm e})^{4}\int_{0}^{1}
(1-\frac{1}{\gamma^{2}_{\rm e}}-\chi ^{2})^{\frac{3}{2}}d\chi\nonumber\\
&&- 2\pi \gamma_{\rm e}(\frac{m_{\rm e}c}{h})^{3}\sqrt{2B{*}}= N_{A}\rho Y_{\rm e}
 \end{eqnarray}

In the case of field-free, for reactions $e^{-}+ p \rightarrow n+ \nu_{\rm e}$
and $n \rightarrow p +e^{-}+ \nu^{-}_{e}$ to take place, there exists the
following inequality among the Fermi momenta of the proton($p_{F}$), the
electron($k_{F}$)and the neutron ($q_{F}$): $p_{F}+ k_{F}\geq q_{F}$
required by momentum conservation near the Fermi surface.  Together with
the charge neutrality condition, the above inequality brings about the
threshold for proton concentration $Y_{\rm p}= n_{\rm p}/(n_{\rm p}+ n_{\rm n})\geq
\frac{1}{9}$= 0.11, this means that, in the field-free case, direct Urca
reactions are strongly suppressed by Pauli blocking in the neutron-rich
nuclear matter formed only by protons, neutrons and electrons.  However,
in a magnetic field $B\gg B_{\rm cr}$, direct Urca reactions are open for
an arbitrary proton concentration ($Y_{\rm e}\leq$ 0.11) due to the fact
that strong magnetic field can alter matter compositions and increase
phase pace for protons which leads to the increase of $Y_{\rm e}$ \citep{lai91}.
Calculations indicate that $E_{\rm F}(\rm e)$ is (39.3$\sim$ 42.2)$(B/B_{\rm cr})
^{\frac{1}{4}}$ MeV corresponding to $Y_{\rm e}\sim$ 0.08 - 0.11 at a given
nuclear density 2.8$\times 10^{14}$g~ cm$^{-3}$.  We assume that direct
Urca reactions must occur in the core of neutron star.  According to
the calculation above, in the range of allowable error ($\leq 5\%$) we
gain an approximate relationship between $E_{\rm F}(\rm e)$ and $B$, which can
be expressed as $E_{\rm F}(\rm e)$=40$(B/B_{\rm cr})^{\frac{1}{4}}$ MeV when $\rho$=
2.8$\times 10^{14}$g~ cm$^{-3}$ and $Y_{\rm e}$ is $\sim$ 0.08- 0.11.

(Cited from the paper:  `Neutron Star Magnetic Field and The Electron Fermi
Energy' Authors:  Gao Z. F., Wang N.,Yuan J.P., et.al., 2010, Prepared)


\begin{thebibliography}{}
\bibitem[Bachcall \& Wolf(1965)]{bac65}
Bachcall J.N.,Wolf R.A.,1965,\pra,140(5B),1452
\bibitem[Baiko \& Yakovlev(1999)]{bai99}
Baiko D.A., Yakovlev D.G.,1999,\aa,342,192-200
\bibitem[Canuto \& Chiu(1971)]{can71}
Canuto V., Chiu H.Y.,1971, Space Sci. Rev.,12,3
\bibitem[Canuto \& Ventura (1977)]{can77}
Canuto V., Ventura J.,1977, Fund. Cosmic Phys.,2,203
\bibitem[Debades et al.(1998)]{deb98}
Debades. B., Somenath. C., Prantick.D. et.al.,arXiv:astro-ph/9804145 v1.
\bibitem[Duncan \& Thompson(1992)]{dun92}
Duncan R.C., Thompson C., 1992, \apj, 392, L9
\bibitem[Gamov \& Schoenberg(1941)]{gam41}
Gamov G., Schoenberg M., 1941, \pra, 59,539
\bibitem[Gao et al(2010)]{gao10}
Gao Z.F., Wang N.,Yuan J.P., et.al., 2010, `Neutron Star
 Magnetic Field and The Electron Fermi Energy ' Submitted.
\bibitem[Kouveliotou et al(1998)]{kou98}
Kouveliotou C., Dieters S., Strohmayer T., et al.,
1998,\nat,393, 235
\bibitem[Kouveliotou et al(1999)]{kou99}
Kouveliotou C., Strohmayer T., Hurley K., et al.,
1999, \apj, 510, L115
\bibitem[Lai \& Shapiro(1991)] {lai91}
Lai Dong., Shapiro, S. L., 1991,\apj, 383, 745-751
\bibitem[Megreghetti(2008)]{meg08}
Megreghetti S., 2008, arXiv: 0804.0250
\bibitem[Paczynski(1992)]{pac92}
Paczynski B., 1992, \actaa, 42, 145
\bibitem[Page et al(2006)]{pag06}
Page D., Geppert U., Weber F., 2006, \nphysa, 777, 497
\bibitem[Peng \& Luo(2007)]{pen06}
Peng Q.-H., Luo Z.-Q., 2006, \cjaa, 6, 248, 253
\bibitem[Peng \& Tong(2007)]{pen07}
Peng, Q.-H., Tong, H., 2007, \mnras, 378, 159
\bibitem[Peng \& Tong(2009)]{pen09}
Peng Qiu He., Tong Hao., arXiv:0911.2066v1 [astro-ph.HE]
11 Nov 2009, $10^{th}$ Symposium on Nuclei in the Cosmos,
27 July-1 August 2008 Mackinac Island, Michigan,USA
\bibitem[Pethick(1992)]{pet92}
Pethick C.J., 1992, Rev. Mod. Phys, 6(4),1133
\bibitem[Shapiro et al(1983)]{sha83}
Shapiro S. L., Teukolsky S. A., 1983, "Black holes,white
drarfs,and neutron stars" John Wiley \& Sons, New York
\bibitem[Thompson \& Duncan(1993)]{tho93}
Thompson C., Duncan R. C., 1993, \apj, 408, 194
\bibitem[Thompson \& Duncan(1995)]{tho95}
Thompson C., Duncan R.C., 1995, \mnras,275, 255
\bibitem[Thompson \& Duncan(1996)]{tho96}
Thompson C., Duncan R.C., 1996, \apj, 473, 322
\bibitem[Yakovlev et al.(2001)]{yak01}
Yakovlev D.G., Kaminker A.D., Gnedin O.Y., et.al.,2001,
\physrep,354,1

\end{thebibliography}
\end{document}